\documentclass[twocolumn,superscriptaddress,showpacs,amsfonts,amsmath,amssymb]{revtex4-1}
\pdfoutput=1

\usepackage{epsfig}
\usepackage{graphicx}
\usepackage{color}
\usepackage{braket}

\begin{document}
%
\title{Geometric driving of two-level quantum systems} 
%
%
%
\author{Zu-Jian Ying}
\affiliation{School of Physical Science and Technology, Lanzhou University, Lanzhou 730000, China}
\affiliation{CNR-SPIN, c/o Universit\`a di Salerno, I-84084 Fisciano (Salerno), Italy}
%
\author{Paola Gentile}
\affiliation{CNR-SPIN, c/o Universit\`a di Salerno, I-84084 Fisciano (Salerno), Italy}
\affiliation{Dipartimento di Fisica ``E. R. Caianiello'',
Universit\`a di Salerno, I-84084 Fisciano (Salerno), Italy}
\author{Jos\'e Pablo Baltan\'as}
\affiliation{Departamento de F\'isica Aplicada II, Universidad de Sevilla, E-41012 Sevilla, Spain}
\author{Diego Frustaglia}
\affiliation{Departamento de F\'isica Aplicada II, Universidad de Sevilla, E-41012 Sevilla, Spain}
\affiliation{%
	Freiburg Institute for Advanced Studies (FRIAS), 
	Albert-Ludwigs Universit\"at Freiburg, 
	D-79104 Freiburg, 
	Germany}
\author{Carmine Ortix}
\affiliation{Institute for Theoretical Physics, Center for Extreme Matter and Emergent Phenomena, Utrecht University, Princetonplein 5, 3584 CC Utrecht, The Netherlands}
\affiliation{Dipartimento di Fisica ``E. R. Caianiello'',
Universit\`a di Salerno, I-84084 Fisciano (Salerno), Italy}

\author{Mario Cuoco}
\affiliation{CNR-SPIN, c/o Universit\`a di Salerno, I-84084 Fisciano (Salerno), Italy}
\affiliation{Dipartimento di Fisica ``E. R. Caianiello'',
Universit\`a di Salerno, I-84084 Fisciano (Salerno), Italy}

\begin{abstract}
We investigate a class of cyclic evolutions for
driven two-level quantum systems (effective spin-1/2) with a particular focus on the geometric characteristics of the driving and their specific imprints on the quantum dynamics. By introducing the concept of geometric field curvature for any field trajectory in the parameter space we are able to unveil underlying patterns in the overall quantum behavior: the knowledge of the field curvature provides a non-standard and fresh access to the interrelation between field and spin trajectories, and the corresponding quantum phases acquired in non-adiabatic cyclic evolutions. In this context, we single out setups in which the driving field curvature can be employed to demonstrate a pure geometric control of the quantum phases. Furthermore, the driving field curvature can be naturally exploited to introduce the geometrical torque and derive a general expression for the total quantum phase acquired in a cycle. Remarkably, such relation allows to access the mechanisms controlling the changeover of the quantum phase across a topological transition and to disentangle the role of the spin and field topological windings.
As for implementations, we discuss a series of physical systems and platforms to demonstrate how the geometric control of the quantum phases can be realized for pendular field drivings. This includes setups based on superconducting islands coupled to a Josephson junction and inversion asymmetric nanochannels with suitably tailored geometric shapes. 
\end{abstract}

\maketitle
\noindent 




\section{Introduction}

A geometric description is often encountered in physics for providing a unifying conceptual framework to fundamental theories, as successfully demonstrated by the geometric reformulation of special relativity and the construction of general relativity. A geometric perspective in quantum mechanics bloomed after the remarkable discovery \cite{berry84,simon83,pancharatnam56} that a cyclic evolution can be marked by a geometric phase for an adiabatically perturbed system. The emerging geometric phase naturally connects with the ubiquitous concept of gauge fields in physics and to the mathematical notion of fiber bundle. The progress along this direction led to the generalization of the geometric phase in degenerate quantum systems \cite{wilczek84} and nonadiabatic cyclic evolutions \cite{aharonov87} considering the connection's property of the projective Hilbert space, which is defined as the set of rays of the Hilbert space. In this context, the geometric phase factor refers to the parallel transport trasformation around a closed curve with respect to the natural connection in the projective Hilbert space as given by the inner product. Starting from these seminal works, the concept of geometric phase has been further developed, setting its relation with the area enclosed by the cyclic trajectory on the corresponding domain of the projective space. This approach has further led to the remarkable observation that there is a nontrivial geometric phase even for classical systems \cite{hannay85,malykin03,bookCLQ}.
Alternative advancements have brought to the construction of the geometric phase in non-cyclic evolution \cite{bhandari88,anandan90,mukunda93} where, for an arbitrary quantum trajectory, it is also possible to show that the integral of the uncertainty of energy with respect to time is independent of the particular Hamiltonian used to transport the quantum system along a given curve in the projective Hilbert space \cite{anandan90}. On a general ground the geometry of quantum states in the Hilbert space is encoded in the quantum metric tensor \cite{Kolodrubetz2017,Provost1980} whose real (i.e. Fubini-Study metric) and imaginary (i.e. Berry curvature) components have been successfully measured in a large variety of engineered quantum platforms.

In the domain of quantum information processing a special position is given to driven two-level systems (TLSs) as a paradigmatic model to describe a large variety of physical systems. Indeed, it was originally used in relation to spins and atomic collisions, and then extended to artificial mesoscopic systems based on semiconducting quantum dots and superconducting circuits. 
A distinct aspect of the quantum TLS is that the two energy levels can exhibit an avoided level crossing when some
external parameters are varied. The physical properties of the two energy eigenstates are typically exchanged
when going from one side of the avoided crossing to the other side. If the external control parameter is varied in time such
that the system crosses the avoided region, a non-adiabatic Landau-Zener transition can occur \cite{landau32,zener32,stueckelberg32,majorana32}.
Along this line, solid-state TLSs are at the center of great attention because they both manifest fundamental
quantum phenomena at a macroscopic scale, and have a great potential to operate as quantum bits (qubits)
in emergent technologies for quantum information processing.

One of the primary goals in quantum information and computation is to implement precise universal gates, because they represent the fundamental building blocks for constructing complex quantum operations. 
A promising approach towards this goal is to use quantum geometric phases which are acquired whenever a quantum system evolves cyclically along a path in the Hilbert space of quantum states. In contrast to dynamical phases, geometric phases depend only on the geometry of the paths executed and are therefore robust to perturbations or certain types of errors, thus offering a significant potential to improve the fidelity of the gate operations \cite{unanyan99,duan01,fuentes02,recati02,solinas03}. Although quantum error correction, error-avoiding, and error-suppression methods \cite{zanardi97,viola99} have been developed to control quantum information against decoherence, the geometric \cite{zanardi99,pachos00} and topological \cite{kitaev97,freedman00} approaches may provide superior paths to stabilize the quantum evolution by encoding its dynamics into global properties rather than on the details of the way it is actually realized. 
For instance, concerning the manipulation of the holonomic phase, the significant advancements and developments of semiconductor based quantum electronics and nanotechnologies led to the manipulation of electronic states through the corresponding spin geometric phase with experimental evidences \cite{NTKKN12,nagasawa13} and the prospect of achieving topological spin engineering \cite{SVLBNNF15,RBSVLNF17}. 
In this framework, the electron spin can be controlled when combining spin-orbit coupling in inversion asymmetric semiconducting nanochannels with non-trivial geometric curvature.  The potential of this union indeed yields augmenting paths for the design of topological states \cite{gentile15,SVLBNNF15,ying16,RBSVLNF17,pandey18,francica19} and spin-transport \cite{frustaglia04,KNvV04,bercioux05,vVKN06,KSN06,Qetal11,nagasawa13}. Such effects have multifold geometrical marks as they can strongly depend on the nanoscale shaping in narrow spin-orbit coupled semiconducting channels which, in turn, act as driving fields with spatially inhomogeneous geometrical torque controlling both the spin-orientation and its spin-phase through non-trivial spin windings \cite{SVLBNNF15,ying16,RBSVLNF17}. 



In this paper, we study two-level quantum systems subject to driving fields that evolve cyclically in a parametric space by introducing the concept of geometric curvature for any given field trajectory. The main goal is to unveil its role in imprinting the overall quantum behavior. We devise quantum TLS setups on which the driving field's curvature can be employed to control the geometric phase and travel the parameter space along paths that keep the dynamical phase constant. This is demonstrated for pendular fields that can be implemented in different solid-state platforms. By exploiting the knowledge of driving field curvatures, we show the path to construct non-adiabatic solutions that well reproduce most of the quantum phases acquired along closed paths in the parameters space. Moreover, we find that the field curvature unveils the mechanism through which driving fields undergoing a topological transition leave a topological imprint in the quantum TLS dynamics and phases \cite{SVLBNNF15,RBSVLNF17}.  
 

As for physical realizations, we devise a series of platforms exploiting the geometrical character of the driving field and demonstrate its potential to engineer the overall quantum phases. These platforms, such as spin-orbit coupled nanochannels with non-trivial geometric shape and voltage-driven superconducting nanostructures, can be mapped onto spin-1/2 systems with a parametric field driving where predictions of the geometrical mark can be assessed. 

The paper is organized as follows. In Sec. II we define the model system, we provide a quantum dynamical construction of near-adiabatic solutions and apply them to the case of a pendular field. Sec. III is devoted to the introduction of the field curvature concept, the emergent geometrical torque and the general consequences on the total quantum phase acquired during the cycle. In Sec. IV we revisit the near-adiabatic solution from a topological perspective of the spin trajectory on the Bloch sphere. Sec. V is devoted to the discussion of the total phase across a topological transition in the parameters space. Finally, in the concluding section we consider possible physical platforms to observe the predicted effects. 






\section{Spin-$\frac{1}{2}$ systems and the adiabatic approximation in the rotating frame: the pendular field case}
\label{NAA}

We start out by considering the quantum evolution of a generic quantum TLS under the action of time-dependent periodic fields which, for simplicity, we take to be coplanar~\cite{footnote-0}. The corresponding Hamiltonian can be then recast in the following form 
\begin{eqnarray}
{\cal H}(t)= B_x(t) \sigma_x + B_y(t) \sigma_y \,,
\label{Hgen}
\end{eqnarray}
where $B_{x,y}(t)$ are the two components of the $T$-periodic field ${\bf B}$ while $\sigma_{x,y}$ are the corresponding Pauli spin-$\frac{1}{2}$ operators. Assuming that at an initial time $t=0$ the system is prepared in an eigenstate of the Hamiltonian, and that the  applied field changes sufficiently slowly during the course of time, one can suppose that the system will remain in an instantaneous (snapshot) eigenstate of ${\mathcal{H}}(t)$ for all $t \in \left[0, T \right]$. This is the content of the well-known adiabatic approximation (AA). Furthermore, the time periodicity of the driving field ensures that at time $t=T$ the system's state verifies $\ket{\psi(T)}=\ket{\psi(0)} e^{i \phi(T)}$, with a total phase that can be split in geometric and dynamical components. Within the AA, the geometric phase corresponds to the usual Berry phase $\gamma_{B}=\int_0^T {\mathcal A}(t) dt$ with $\mathcal{A}= \bra{\psi} i \partial_t \ket{\psi}$ the Berry connection. In this context, it can be also shown that the geometric phase is proportional to the solid angle $\Omega$ gathered by $\ket{\psi(t)}$ in the Bloch's sphere after one period $T$ (interestingly, this still holds in the case of non-adiabatic dynamics). The dynamical phase is given by $d=- \int_0^T  E(t) dt / \hbar$, where $E(t)$ is the snapshot eigenenergy of the system. For the Hamiltonians class of Eq.~(\ref{Hgen}) the dynamical phase is simply $d= -s \int_0^T |{\bf B}(t)|/\hbar$ where $s=\pm 1$ labels the two non-degenerate quantum levels. Moreover, by  
choosing the gauge in which the snapshot eigenstates read $\ket{\psi(t)}=\left[1, s \exp{(i \vartheta(t))} \right]/\sqrt{2}$,
 the Berry connection can be written as $\mathcal{A}(t)= - \partial_t \vartheta(t) / 2$ where $\vartheta(t)=\arctan{\left[B_y(t)/B_x(t)\right]}$. 

\begin{figure*}
\includegraphics[width=0.95\textwidth]{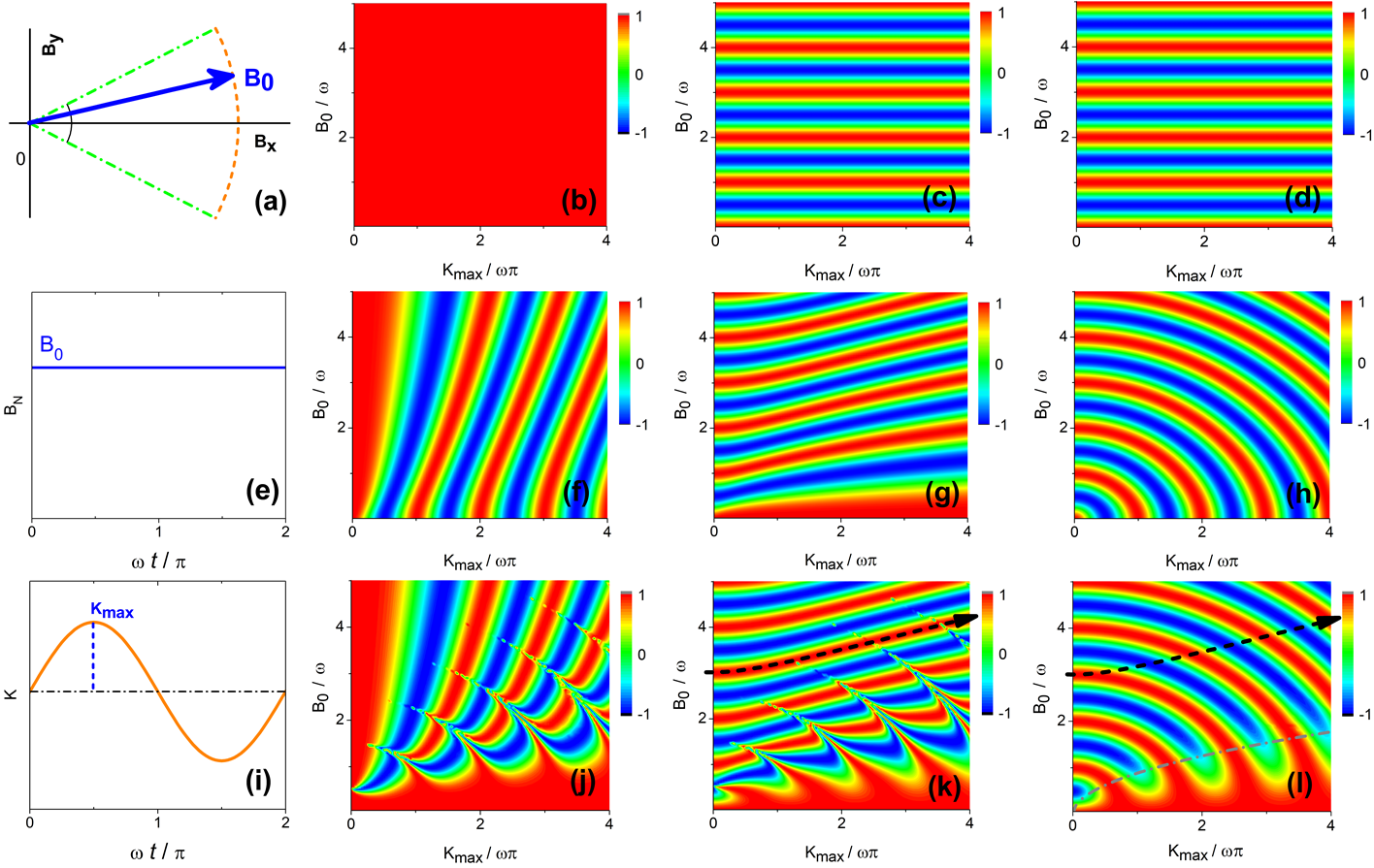}
\caption{(a) Schematic illustration of the pendular field trajectory (orange dots) in the $(B_x,B_y)$ plane. 
Contour map of the cosine of the (b) quantum geometric, (c) dynamical and (d) total phase for the adiabatic solution (AA) corresponding to a spin state that instantaneously follows the direction of the applied field, as a function of the pendular field amplitude $B_0$ and the maximum of the field curvature $K_{\rm max}/\pi$, respectively. (e) Amplitude of the applied field vs time. For the pendular driving field the amplitude $B_0$ is constant in time. 
Contour map of the cosine of the (f) quantum geometric, (g) dynamical and (h) total phase for the adiabatic solution in the rotating frame (AARF) as a function of the pendular field amplitude $B_0$ and the maximum of the field curvature $K_{\rm max}/\omega \pi$, respectively. (i) Time dependent evolution of the field curvature $K(t)$ showing a sinusoidal profile. (j) Quantum geometric, (k) dynamical and (l) total phase for the exact solution of the two-level driven system, respectively. As one can notice, the AARF solution with the spin following the effective field in the rotated frame captures the main features of the quantum geometric, dynamical and total phases. The dotted arrow in panels (k) and (l) indicates a representative path in the parameters space with constant dynamical phase, such that the corresponding variation of the total quantum phase (l) is uniquely due to a geometric modification of the accumulated phase in a cycle. In the region below the long-short dotted line the AARF solution fails and the geometric, dynamical and total phases deviates significantly from those obtained by means of the full solution because the amplitude of the total field in the rotating frame is larger than the corresponding curvature (see Sec. III). We assume $\hbar=1$ in all the panels.
}
\label{fig:fig1}
\end{figure*}

For illustration, it is instructive to consider how these concepts apply to a specific case. Figure 1(a) depicts a pendular driving field of constant magnitude $B_0$ oscillating with frequency $\omega =2\pi/T$ and components
\begin{eqnarray}
\label{pend-1a}
B_x(t)&=&B_0 \cos \left[ \vartheta_0 \cos(\omega t) \right], \\
\label{pend-1b}
B_y(t)&=&B_0 \sin \left[ \vartheta_0 \cos(\omega t) \right] ,
\end{eqnarray}
where $\vartheta(t) = \vartheta_0 \cos(\omega t)$ is the polar angle. By following the above definitions we find a vanishing Berry phase, $\gamma_{B}=0$, and a dynamical phase $d = -s B_0T/\hbar$, as shown in Figs. \ref{fig:fig1}(b), \ref{fig:fig1}(c), and \ref{fig:fig1}(d). This elementary response, however, is dramatically enriched out of the AA when considering a solution which is non-adiabatic and includes curvature effects of the driving field.


Generally speaking, the AA is an appropriate description of the dynamics when the driving period $T$ is much larger than the characteristic relaxation time $\tau(t)=\hbar/|{\bf B}(t)|$ corresponding to the transition between the two quantum levels of the system. 
As a result, corrections to the AA can be  defined perturbatively in the small frequency parameter $1/T$ and, at the first order, yield the so-called near-adiabatic approximation. Instead of employing the latter, we will now define an adiabatic approximation in a particular rotating frame, inspired by the idea put forward by Berry of performing  a series of unitary transformations to the time-dependent Schr\"odinger equation \cite{B87}.

Let us consider the time-depedent Schr\"odinger equation for our spinorial wavefunction:
\begin{equation}
i \hbar \partial_t |\psi(t)\rangle= {\mathcal H}(t) |\psi(t)\rangle 
\end{equation} 
and recall that, using the quantities defined above, the time-dependent Hamiltonian can be recast in the form 
\begin{equation} 
 {\mathcal H}(t)= |{\bf B}(t)|\left[ \cos{\vartheta(t)} \sigma_x +  \sin{\vartheta(t)} \sigma_y \right].
 \label{eq:hamin}
 \end{equation}

\noindent Next, we perform an ${\mathcal{SU}}(2)$ transformation of the Hamiltonian such that the spin operators are instantaneously  aligned with the field amplitude while preserving the structure of the Hamiltonian operator, i.e., its anticommutation with one generator of the Clifford algebra. 
By recalling that the time-dependent Schr\"odinger equation for the transformed wavefunction $|\psi_R(t)\rangle= U^{\dagger}(t) |\psi(t)\rangle$ reads as 
\begin{equation}
i \hbar \partial_t |\psi_R(t)\rangle= \left[ U^{\dagger}(t) {\mathcal H}(t) U(t) - i \hbar U^{\dagger}(t) \partial_t U(t)  \right] |\psi_R(t)\rangle, 
\end{equation} 
we find that the required ${\mathcal{SU}}(2)$ transformation of the Hamiltonian simply reads $U(t)=\exp{\left[-i \vartheta(t) \sigma_z / 2\right]}$. 
Consequently, the time-dependent Schr\"odinger equation for the rotated wavefunction is given by
\begin{equation}
i \hbar \partial_t |\psi_R(t)\rangle= \left[|{\bf B}(t)| \sigma_x +  \dfrac{\hbar K(t)}{2} \sigma_z \right] |\psi_R(t)\rangle
\label{eq:transf_hamiltonian}  
\end{equation} 
where we have introduced $K(t)=-\partial_t \vartheta(t)$ for later convenience. 
Two remarks are in order here. First, the fact that a rotation of the wavefunction yields a different time-dependence in the Hamiltonian -- it also involves the velocity of the driving fields --  
allows us to establish an ``instantaneous" criterion for the validity of the quantum adiabatic approximation. 
In fact, the latter will be accurate as long as $\text{min}|{\bf B}(t)| \gg \text{max} |\hbar K(t)|$, so that the quantum evolution of the system is not susceptible to the instantaneous rotation of the Hamiltonian. Second, we can now define an adiabatic approximation in the rotating frame (AARF) by demanding
the rotated wavefuntion to be a snapshot eigenstate of the rotated
Hamiltonian written above, $\ket{\psi_{RA}(t)}$. This, in turn, yields the non-adiabatic wavefunctions
$\ket{\widetilde{\psi}(t)}=\widetilde{U}(t)\ket{\psi_{RA}(t)}$ in the
laboratory frame, where $\widetilde{U}(t)=\mathcal{U}(1) U(t)$ includes
an additional $\mathcal{U}(1)=\exp{\left[-i \vartheta(t)/ 2\right]}$
unitary transformation guaranteeing that $\ket{\widetilde{\psi}(t)}$
is periodic. Hence, we can compute the non-adiabatic Aharonov-Anandan
geometric phase $\gamma=\int_{0}^{T}\bra{\widetilde{\psi}(t)} i
\partial_t \ket{\widetilde{\psi}(t)} dt$ as well as the non-adiabatic
dynamical phase in a straightforward manner. In fact the two phases
take the simple form
\begin{eqnarray}
\gamma&=& \int_{0}^T \braket{\widetilde{U}^{\dagger}(t) i \partial_t \widetilde{U}(t)} dt + \gamma_B \\ 
d&=& - \dfrac{1}{\hbar} \int_{0}^{T} \braket{\widetilde{U}^{\dagger}(t) \mathcal{H}(t) \widetilde{U}(t)} dt \nonumber \\
&=& - \dfrac{1}{\hbar} \int_{0}^{T} |{\bf B}(t)| \braket{\sigma_x} dt
\end{eqnarray}

In the equations above, the geometric phase consists of two terms. The first term corresponds to the expectation value over the snapshot eigenstates $\ket{\psi_{RA}(t)}$ of the composed unitary transformation, while the second term corresponds to the Berry phase $\gamma_B=\int_{0}^{T}\bra{\psi_{RA}(t)} i \partial_t \ket{\psi_{RA}(t)} dt$, which identically vanishes. The dynamical phase simply corresponds to the expectation value of the Hamiltonian in Eq.~(\ref{eq:hamin}) over the non-adiabatic wavefunctions $\ket{\widetilde{\psi}(t)}$, which in terms of the adiabatic $\ket{\psi_{RA}(t)}$ can be written as the spin expectation value $\braket{\sigma_x}$. Finally, by using the conventional expression for the snapshot eigenstates $\ket{\psi_R(t)}$,  we end up with the following expression for the two quantum phases
\begin{eqnarray}
\gamma&=&\dfrac{1}{2} \int_0^T K(t) dt -  \dfrac{s}{2} \int_0^T \dfrac{\hbar K(t)^2}{\sqrt{4 |{\bf B}(t)|^2 + \hbar^2 K(t)^2}} dt \, , \label{geomph-1}\\ 
d&=& -  \dfrac{s}{\hbar} \int_0^T  \dfrac{2 |{\bf B} (t)|^2}{\sqrt{4 |{\bf B} (t)|^2 + \hbar^2 K(t)^2}} dt. \label{dynph-1} 
\end{eqnarray}
It is instructive to examine  the approximate dynamics introduced above from a {\it geometric} viewpoint.
For this purpose it is convenient to employ a moving reference frame with a time-dependent basis spanned by two unit vectors, $\hat{\cal N}(t)$ and $\hat{\cal T}(t)$, that are defined at any given time $t$ in the applied field's space. In a similar fashion, one can also define the local Pauli matrices projected along $\hat{\cal N}(t)$ and $\hat{\cal T}(t)$ in the moving frame as
$\sigma_{\cal N}(t) = \boldsymbol{\sigma} \cdot \hat{\cal N}(t)$ and $\sigma_{\cal T}(t) = \boldsymbol{\sigma} \cdot \hat{\cal T}(t)$. The choice of the reference frame is made in such a way to have the applied field always collinear to one direction [e.g. $\hat{\cal N}(t)$]. 
Hence, as it is commonly done for the case of a generic curvilinear profile in two dimensions, one can conveniently set $\hat{\cal N}(t)$ and $\hat{\cal T}(t)$ as the normal and tangential directions of the effective field trajectory and employ the polar angle
$\vartheta(t)$ to express them in parametric form as $\hat{\cal N}(t) = \left\{\cos{\vartheta(t)},\sin{\vartheta(t)} ,0 \right\}$, and  $\hat{\cal T}(t) = \left\{\sin{\vartheta(t)} , -\cos{\vartheta(t)},0 \right\}$. By using the Frenet-Serret (FS) equations~\cite{frenet-serret}, it is then possible to connect the variation of the normal component with the tangential one through the relation $\partial_t \hat{\cal N}(t) = K(t) \hat{\cal T}(t)$, which defines {\it the local curvature} $K(t)$ {\it of the field trajectory} (the field curvature in what follows) in the moving frame. This directly implies that the polar angle $\vartheta(t)$ and the local curvature are related via $\partial_t \vartheta(t) = -K(t)$, which in turn endows the effective field $z$-component $K(t)$ introduced in Eq.~(\ref{eq:transf_hamiltonian}) with a precise geometrical meaning. Put in different words, the local field curvature is equivalent to an extra field component along the $z$-direction in the rotating frame. We will elaborate on this connection in the following Section.

\begin{figure*}
\includegraphics[width=0.95\textwidth]{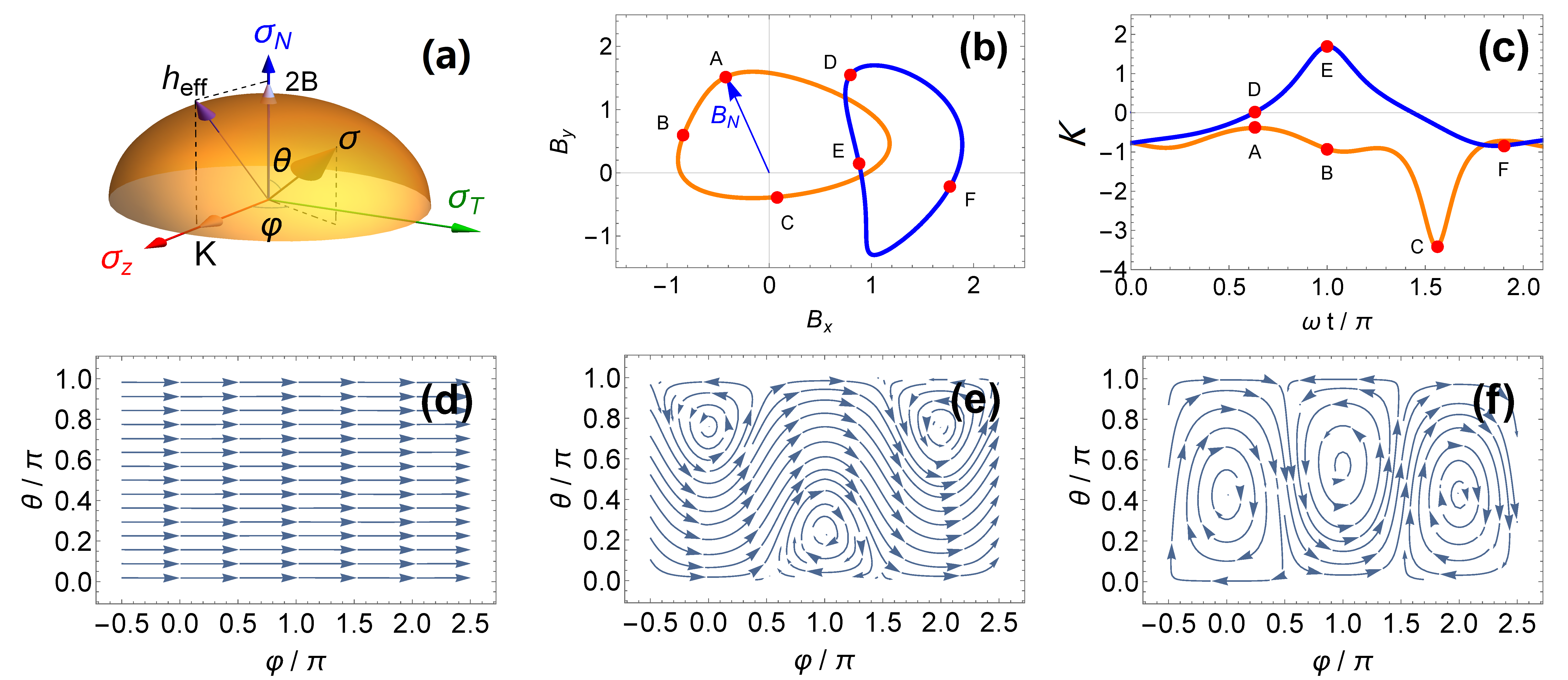}
\caption{Schematics of (a) Frenet-Serret-Bloch sphere with the effective torque field $\bf{h}_{\text{eff}}$, (b) parametric profile of two representatives field trajectories with different winding number and (c) the corresponding curvatures $K(t)$. 
Panels (d),(e) and (f) denote a time snapshot of the average spin orientation on the Frenet-Serret-Bloch sphere in the $(\theta,\phi)$ plane corresponding to points D, B and C in panel (c), respectively.}
\label{fig:fig2}
\end{figure*}   

We observe that in the selected rotating frame the Hamiltonian can be expressed as
\begin{eqnarray*}
{\cal H}(t)= |{\bf B}(t)| \sigma_{\cal N}(t) \, ,
\end{eqnarray*}
with $\sigma_{\cal N}$ reading
\begin{eqnarray}
\sigma_{\cal N}(t)=[f_x(t) \sigma_x +f_y(t) \sigma_y] \,.
\label{Eq:sigmaN}
\end{eqnarray}
Here, $f_x(t)=\frac{B_x(t)}{|{\bf B}(t)|}$ and $f_y(t)=\frac{B_y(t)}{|{\bf B}(t)|}$ are the projections of the spin components along the $x$ and $y$ axes in the lab reference frame, respectively. By using Eq.~(\ref{Eq:sigmaN}) and the relation between the polar angle and the curvature, one can immediately deduce the expression of the effective field curvature in terms of the field components as
\begin{eqnarray}
K(t)=-[f_x(t) \partial_t f_y(t) -f_y(t) \partial_t f_x(t) ] \,.
\label{Eq:ktgen}
\end{eqnarray}
As a first observation, by virtue of the FS geometric representation, we find that the integral of the curvature over a period is an integer $n_{K}$ modulo $2 \pi$, namely
\begin{eqnarray}
{\frac{1}{2 \pi} \int_{0}^{T} K(t) dt = n_{K}} \,.
\label{Eq:windingK}
\end{eqnarray}
Indeed, it is equivalent to the winding of the applied field and thus provides information on the topological character of the driven quantum system with respect to the field trajectory in the time space. In Fig. \ref{fig:fig2}(b) we show two generic field trajectories associated with either zero or non-vanishing windings. According to Eq.~(\ref{Eq:ktgen}), one can directly determine the corresponding evolution of the geometric curvature $K(t)$ [see Fig. \ref{fig:fig2}(c)]. 
As expected, for the zero-winding field trajectory the curvature changes its sign, while it has a unique sign for the case of a field that winds around the origin. We also notice that the amplitude of the curvature is generally non-uniform in time and it can get enhanced at special points of the trajectory. This can be observed, for instance, in the positions E and C of the trajectories in Fig. \ref{fig:fig2}(b). Alternatively, there can be positions along the time evolution where the curvature is small or vanishes as it occurs at the points A,F and D in Fig. \ref{fig:fig2}(c), respectively.



Back to the pendular driving introduced by Eqs. (\ref{pend-1a}) and (\ref{pend-1b}) and depicted in Fig. \ref{fig:fig1}(a), we find that the angular amplitude reads $\vartheta_0=K_{\rm max}/{\omega}$, where $K_{\rm max}$ is the maximum value taken by the curvature $K(t)=\vartheta_0 \omega \sin(\omega t)$, such that
\begin{eqnarray}
\label{pend-2a}
B_x(t)&=&B_0 \cos \left[ \frac{K_{\rm max}}{\omega} \cos(\omega t) \right], \\
\label{pend-2b}
B_y(t)&=&B_0 \sin \left[ \frac{K_{\rm max}}{\omega} \cos(\omega t) \right].
\end{eqnarray}
As expected for a pendular field with trivial topology, we notice that the winding $n_K$ defined in Eq.~(\ref{Eq:windingK}) vanishes. Still, this does not prevent the system to develop a complex dynamics in non-adiabatic conditions. This can be seen by evaluating the geometric and dynamic phases arising from the AARF given in Eqs.~(\ref{geomph-1}) and (\ref{dynph-1}), the solution of which are elliptic integrals depicted in Figs. \ref{fig:fig1}(f) and \ref{fig:fig1}(g) as a function of the field's strength $B_0$ and the curvature's amplitude $K_{\rm max}$ (in units of $\omega$). 
There we find that the geometric phase, Fig. \ref{fig:fig1}(f), displays a series of wavefronts mainly controlled by $K_{\rm max}$ with a drift as a function of $B_0$. This stands in sharp contrast to the case of adiabatic evolution with vanishing Berry phase for the spin solution that istantaneously follows the field trajectory, Fig. \ref{fig:fig1}(b). As for the dynamical phase [Fig. \ref{fig:fig1}(g)], it develops wavefronts as a function of $B_0$ similar to those found in the AA [Fig. \ref{fig:fig1}(c)], except that for the AARF it exhibits a drift as a function of $K_{\rm max}$. Due to the geometric phase contribution, the total phase [Fig. \ref{fig:fig1}(h)] now displays a pattern of radial wavefronts differing significantly from the standard adiabatic case as reported in Fig. \ref{fig:fig1}(d). 

In Figs. \ref{fig:fig1}(j)--(l) we show the exact solutions for the geometric, dynamic and total phases by solving the full dynamics of the two-level system under the pendular driving described by Eqs.~(\ref{pend-2a})-(\ref{pend-2b}) employing both the Floquet approach and the discretization of the time dependent differential equations. By comparison with Figs. \ref{fig:fig1} (f)--(h), we find that the AARF captures the main features of the geometric and dynamic phases except for the set of localized dynamical degeneracies (vanishing dynamical phases) emerging under strong driving (coinciding with Rabi resonances for small $\vartheta_0=K_{\rm max}/\omega$). As for the total phase [see Figs. \ref{fig:fig1}(h) and \ref{fig:fig1}(l)], the AARF also captures the overall behavior very well thanks to the exact cancellation of dynamical-degeneracy contributions present in geometric and dynamical phases as it has been also reported in \cite{SVLBNNF15} and \cite{RBSVLNF17} for circular field drivings.  

The above example is very instructive as it already illustrates the role played by the field's curvature in the control of the two-level dynamics. In the following sections we provide another perspective of our approach by discussing further geometric and topological aspects.

\section{Field driven curvature and geometrical torque}

While it is intuitive to single out the topological aspect of the
curvature or winding of the applied field, it is less obvious to track
the meaning and the role of the instantaneous amplitude of the
curvature $K(t)$ at any given position along the parametric
evolution. We aim to show that, indeed, the value of the curvature
carries fundamental information for predicting the overall behavior of
the quantum TLS, and that it plays a role which is beyond
its topological intrinsic character. In particular, some of the
results discussed in this Section apply to any parametric dependence
of the applied field, including the possibility of non-periodic
trajectories.

To start, we recall that the time evolution of a generic spin state
$|\psi (t)\rangle$ is described by the Schr\"{o}dinger equation $i
\hbar \partial_t |\psi (t)\rangle= {\cal H}(t) |\psi (t)\rangle$. Let
us then consider the spin orientation for the state $|\psi (t)\rangle$
defined by the corresponding expectation value of the spin operators in the
FS reference frame, i.e.  $\langle \psi
|\boldsymbol{\sigma} |\psi\rangle= \langle \boldsymbol{\sigma}
\rangle=\{\langle {\sigma}_{\cal T} \rangle,\langle {\sigma}_{\cal N}
\rangle,\langle {\sigma}_z \rangle \}$, where we drop the
time-dependence of the expectation values here and in the following paragraphs
for convenience.  Taking into account both the FS and the
Schr\"{o}dinger equations, one immediately arrives to:
\begin{eqnarray}
\partial_t \langle \boldsymbol{\sigma} \rangle = i \hbar^{-1}\langle [{\cal H}(t), \boldsymbol{\sigma}]\rangle + \langle \partial_t \boldsymbol{\sigma} \rangle \,
\label{eq:deriv}
\end{eqnarray}
\noindent with $[A,B]$ denoting the commutator of $A$ and
$B$. Hence, by considering that $[{\cal H}(t),{\sigma}_{\cal N}]=0$, $[{\cal
    H}(t),{\sigma}_{\cal T}]=-2 i |{\bf B}(t)| {\sigma}_z$, $[{\cal
    H}(t),{\sigma}_z]=2 i |{\bf B}(t)| {\sigma}_{\cal T}$, and $\partial_t
\sigma_z=0$, it follows:
\begin{eqnarray}
\partial_t \langle \sigma_{\cal N} \rangle &=& K(t) \langle \sigma_{\cal T} \rangle \nonumber \\
\partial_t \langle \sigma_{\cal T} \rangle &=& 2 \hbar^{-1} |{\bf B}(t)| \langle \sigma_z \rangle - K(t) \langle \sigma_{\cal N} \rangle \nonumber \nonumber \\
 \partial_t \langle \sigma_z \rangle &=& 2 \hbar^{-1} |{\bf B}(t)| \langle \sigma_{\cal T}\rangle. 
\label{eq:gyro}
\end{eqnarray}
These relations can be rearranged in a compact gyroscope-like form by
introducing an effective time dependent field
${\bf h}_{\text{eff}}(t)=\{0,2 \hbar^{-1} |{\bf B}(t)|,
K(t)\}$ in the space spanned by the the spin components $\langle
\boldsymbol{\sigma} \rangle=\{\langle \sigma_{\cal T} \rangle,\langle
           \sigma_{\cal N} \rangle,\langle {\sigma}_z \rangle \}$. The
           ensuing gyroscope equation reads as
\begin{eqnarray}
  \partial_t \langle \boldsymbol{\sigma} \rangle= {\bf h}_{\text{eff}}(t) \times \langle \boldsymbol{\sigma} \rangle \,.
\label{eq:gyro1}
\end{eqnarray}
Since the time derivative of the spin vector is perpendicular to $\boldsymbol{\sigma}$, it directly follows that the amplitude of the local spin component $\langle \boldsymbol{\sigma} \rangle^2$ is constant along the parametric trajectory, i.e. 
$\partial_t \left(\langle \boldsymbol{\sigma}\rangle \cdot \langle \boldsymbol{\sigma}\rangle \right) = 0$.

The resulting field ${\bf h}_{\text{eff}}(t)$ in the moving
frame is made of two components [Fig. 1(a)]. One points along $\hat{{\cal
  N}}$ and it depends only on the amplitude of the applied field
$|{\bf B}(t)|$. The second one is parallel to the $z$ direction in
the spin space and it has a pure geometrical character in the sense
that it is uniquely linked to the change of orientation of the applied
field through the field curvature $K(t)$. By
construction, then, ${\bf h}_{\text{eff}}(t)$ has a time
evolution that is confined in a plane within the rotating spin
reference frame [Fig. 1(a)], independently of the form of the applied
field in the parametric space. We observe that any orientation change
of the driving field leads to a non-trivial component of
${\bf h}_{\text{eff}}(t)$ along the $z$-direction which is
perpendicular to the plane of the applied field. This is also a
general aspect of ${\bf h}_{\text{eff}}(t)$ and it occurs
independently of the topological character of the applied field, that
is, whether or not the field has a non-vanishing winding regarding its
evolution in the parameters space.
A simple scenario now emerges: in the rotating frame, the spin
evolves in time
according to Eq.~(\ref{eq:gyro1}), subject to a planar effective field to which the field curvature contributes by providing a
geometrical component that is perpendicular to the plane of the
applied field. This sheds new light on the dynamical approach
introduced in Section \ref{NAA}, with
${\bf h}_{\text{eff}}(t)$ being in clear correspondence with
the effective field defined by Eq.~(\ref{eq:transf_hamiltonian}).




It is also convenient to unveil the geometrical and topological aspects encoded in the geometrical and dynamical phases introduced in Section \ref{NAA}. To this end, we note that the wave-function $|\psi (t)\rangle$ can be generally expressed 
in the form
\begin{equation*}
|\psi (t) \rangle=\left( 
\begin{array}{c}
\exp[i f(t)/2]\, \exp[i \theta_{\Uparrow}(t)] A_{\Uparrow}(t) \  \\ 
\exp[-i f(t)/2]\, \exp[i \theta_{\Downarrow}(t)] A_{\Downarrow}(t)%
\end{array}%
\right)
\end{equation*}%
\noindent where $f(t)=\int_{0}^{t} K(\bar{t}) d\bar{t}$, and $\{A_{\Uparrow},A_{\Downarrow}\}$ are real. This structure for $|\psi(t) \rangle$ is 
convenient because the expectation values of  
the local spin $\langle \boldsymbol{\sigma} \rangle$ in the FS reference frame can be linked to the components of the wave-function through the following relations:
\begin{eqnarray}
\tan[\theta_{\Uparrow}-\theta_{\Downarrow}]=\frac{\langle \sigma_{\cal T} \rangle}{\langle \sigma_{\cal N} \rangle} \label{wind} \\
A_{\Uparrow}^2-A_{\Downarrow}^2=\langle {\sigma}_z \rangle \label{sz}
\end{eqnarray}
In addition, the integral of the curvature over a period is a multiple of an integer [Eq.~(\ref{Eq:windingK})].
Interestingly, after a period $T$, the phase difference $(\theta_{\Uparrow}-\theta_{\Downarrow})$ acquires a shift $2\pi n_{\cal N T}$, with $n_{\cal{N}\,\cal{T}}$ being the winding number associated with the normal and tangential spin components:
\begin{eqnarray*}
n_{\cal{N}\,\cal{T}}=\frac{1}{2 \pi} \int_{0}^{T} q_{\cal NT}(t) dt \,.
\end{eqnarray*}
Here, $q_{\cal NT}(t)=\frac{\left[\langle \sigma_{\cal N} \rangle \partial_t \langle \sigma_{\cal T} \rangle -\langle \sigma_{\cal T} \rangle \partial_t \langle \sigma_{\cal N} \rangle \right]}{\left[\langle \sigma_{\cal T} \rangle^2 + 
\langle \sigma_{\cal N} \rangle^2 \right]}$, in analogy with the curvature of the applied field, may be naturally understood as the curvature of the normal and tangential spin components with respect to the binormal direction in the parametric space.

Furthermore, one can show that
\begin{equation*}
|\tilde{\psi} (t) \rangle= \left( 
\begin{array}{c}
A_{\Uparrow}(t) \  \\ 
\exp[i f(t)]\, \exp[-i (\theta_{\Uparrow}(t)-\theta_{\Downarrow}(t))] A_{\Downarrow}(t)%
\end{array}%
\right) 
\end{equation*}
verifies $|\tilde{\psi} (0) \rangle=|\tilde{\psi} (T) \rangle$ which,
according to Aharonov and Anandan, allows us to compute the geometric phase as
\begin{eqnarray}
&&\gamma=\int_{0}^{T} \frac{\langle \tilde{\psi}| i\partial_t |\tilde{\psi}\rangle}{\langle \psi|\psi \rangle} dt= \label{eq:gAA}\\
&&\pi \left( n_K+ n_{\cal{N}\,\cal{T}} -\frac{1}{2 \pi} \int_0^T \langle {\sigma}_z \rangle [K(t) + q_{\cal NT}(t)] dt \right) \nonumber \,\\
\end{eqnarray} 
with the dynamical phase given by
\begin{equation}
d=-\frac{1}{\hbar}\int_{0}^{T} \frac{\langle \psi| {\cal H} |\psi\rangle}{\langle \psi| \psi \rangle} dt=-\frac{1}{\hbar}\int_0^T  |{\bf B}(t)| \langle \sigma_{\cal N} \rangle dt \,.
\label{eq:dynamic1}
\end{equation}

We observe that the geometrical and dynamical phases depend on both
the curvature of the applied field $K(t)$ and the curvature of the
normal and tangential spin components, $q_{\cal NT}(t)$, as well as on the
components of the spin orientation vector and their time derivatives
[via $q_{\cal NT}(t)$], which in turn depend on the amplitude of the spin components
themselves via Eqs.~(\ref{eq:gyro}). This allows one to end up with a fundamental expression for the geometric phase which explicitly shows its interrelation with the dynamical phase and with the field and spin winding numbers as
\begin{eqnarray}
\gamma=-d - \frac{1}{\hbar} \int_0^T \frac{|{\bf B}(t)|
\langle \sigma_{\cal N} \rangle}{\langle \sigma_{\cal N} \rangle^2+\langle \sigma_{\cal T} \rangle^2} dt  + \pi [n_{\cal{N}\,\cal{T}}+ n_{K}] \,.
\label{eq:geometric1}
\end{eqnarray}
Moreover, by reinserting Eqs.~(\ref{eq:gyro}) in
Eq.~(\ref{eq:geometric1}), we obtain for the total phase
\begin{eqnarray}
\phi_{tot}=\gamma+d= - \frac{1}{\hbar} \int_0^T \frac{|{\bf B}(t)|
\langle \sigma_{\cal N} \rangle}{1+\langle {\sigma}_z \rangle} dt,
\label{eq:totalphase}
\end{eqnarray}
which shows that $\phi_{tot}$ is independent of the spin and field
curvatures. This is one of the central results of the manuscript: for a given
cyclic evolution in the parametric space, the total phase acquired by
the quantum state does not depend explicitly  on the velocity of the average spin
components. 
Remarkably, the integrand only differs from that of the
dynamical phase, Eq.~(\ref{eq:dynamic1}), by a factor that depends on the component of the spin
which is perpendicular to the plane of the applied field. In principle, the regularity of the integrand in
Eq.~(\ref{eq:totalphase}) might be compromised by the presence of this
factor only if $1+\langle \sigma_z \rangle \rightarrow 0$, which
corresponds to the spin passing through the south pole on the
Frenet-Serret-Bloch (FSB) sphere. However, a closer look at this case evidences that
the integrand is in fact regular everywhere, so that, {\it on a
  general ground and independently of the form of the driving field},
one does expect a {\it smooth} evolution of the total phase in the
parametric field space. This is shown by first noticing that, for
the specific times $t^{*}$ when $1+\langle \sigma_z \rangle
\rightarrow 0$, $\langle \sigma_{\cal N} \rangle$ also vanishes, which
demands a detailed evaluation of the limit.  A Taylor expansion around
$t^{*}$ of both the numerator and denominator of the integrand gives,
up to zeroth order in $t\rightarrow t^{*}$
\begin{eqnarray}
\frac{|{\bf B}(t^{*})|
\langle \sigma_{\cal N} \rangle}{1+\langle {\sigma}_z \rangle} \sim \frac{|{\bf B}(t^{*})| K(t^{*})\langle \sigma_{\cal T} \rangle_{t^{*}}}
{2 |{\bf B}(t^{*})| \langle \sigma_{\cal T} \rangle_{t^{*}}}=\frac{1}{2} K(t^{*}) \,.
\end{eqnarray}
Since the curvature has a smooth behavior in time, we do not expect 
a singular changeover of the total phase in the parameters space,
even in this critical case.


It is worth to note that, although the amplitude of the tangential
spin-component does not explicitly appear in the expression for
$\phi_{tot}$, it implicitly affects the total phase since the total
amplitude of the spin is a constant of motion, hence $\langle
{\sigma}_{\cal T} \rangle$
plays a role through this constraint. Finally, it
is apparent that significant variations of the total phase may be
expected since spin trajectories may lead to cancellations or
amplifications of the integrand function.

\section{AARF revisited and beyond}

It is instructive to consider the resulting geometric and dynamical phases for a spin trajectory such as the spin orientation is always parallel to the field ${\bf h}_{\text{eff}}(t)$.
For such configurations, one has that $
\langle\sigma_{\cal N} \rangle\equiv {h}_{\cal N}$ and $
\langle\sigma_z\rangle\equiv {h}_{z}$, where ${h}_{\cal N}(t)=\frac{2 |{\bf B}(t)|}{|{\bf h}_{\text{eff}}(t)|}$ and ${h}_{z}(t)=\frac{K(t)}{|{\bf h}_{\text{eff}}(t)|}$ are the projections of ${\bf h}_{\text{eff}}(t)$ along the ${\cal N}$ and $z$ directions in the spin components space. By replacing these expressions in the relations for the geometric and dynamical phases, one finds that 
%
\begin{eqnarray}
\gamma &&=\pi \left(n_K-\frac{1}{2 \pi} \int \frac{\hbar K(t)^2}{|{\bf{h}}_{\text{eff}}(t)|} dt \right) 
\label{NAA-1} \\
d&&=- \int \frac{2 \hbar |{\bf B}(t)|^2}{|{\bf{h}}_{\text{eff}}(t)|} dt 
\label{NAA-2} \\
\phi_{tot}&&=\pi n_K+\int \frac{|{\bf{h}}_{\text{eff}}(t)|}{2 \hbar} dt. 
\label{NAA-3}
\end{eqnarray}
Again, this is in complete agreement with the results obtained within the
AARF discussed in Section \ref{NAA}.
%
%

In order to further comprehend the consequences of the field curvature on the spin trajectory it is convenient to express the torque equation in spherical coordinates in the FSB reference frame [see Fig. 2(a)].
Then, a point on the sphere identifies the average spin orientation at a given time position $t^{*}$ through the angles $\{\theta(t^*),\varphi(t^*)\}$ [Fig. 2(a)]. The average spin components can be written as 
\begin{eqnarray*}
\langle \sigma_{\cal N}(t)\rangle &=& \cos [\theta] \\
\langle \sigma_{z}(t)\rangle &=& \sin [\theta ]\cos [\varphi ] \\
\langle \sigma_{\cal T}(t)\rangle &=& \sin [\theta ]\sin [\varphi ] \, ,
\end{eqnarray*}
where the spin $\sigma$ is assumed to have an amplitude equal to one.
The torque equations (\ref{eq:gyro}) reduce to two independent equations for the derivative of the coordinates $\{\theta(t),\varphi(t)\}$ 
\begin{eqnarray}
\overset{\cdot }{\theta } &=&K(t)\sin[\varphi] \nonumber \\  
\overset{\cdot }{\varphi } &=&2 |{\bf B}(t)|+K(t)\cos[\varphi ]\frac{1}{\tan
  [\theta ]} .
\label{eq:gyro_spherical}
\end{eqnarray}
By assuming that the curvature is non singular along the time trajectory, one observes that the torque vanishes (i.e., that
$\overset{\cdot }{\theta}=\overset{\cdot }{\varphi}=0$) at those
points $P_{1,2}(t)$ on the FSB sphere such that
$\varphi=\overline{\varphi }_{1,2} =0$ or $\pi$ and
$\theta=\overline{\theta }_{1,2}(t) =\mathrm{arc}\cot[\pm \frac{2
  |{\bf B}(t)|}{K(t)}]$. It is worth pointing out that, independently
of the geometric properties of the field trajectory, these points lie
on the line defined by intersection of the FSB sphere and the ${\cal
  N}$-$z$ plane, along which they move with a velocity

\begin{equation}
v_{\overline{\theta}_{1,2}}(t)= \overset{\cdot }{\overline{\theta }}_{1,2}(t)=\pm \frac{2 [-|{\bf B}(t)|^{'} K(t)+|{\bf B}(t)|  K^{'}(t) ]}{4 |{\bf B}(t)|^{2}+K(t)^2} 
\label{eq:speedvortex},
\end{equation}
which is strongly connected to the time evolution of the applied field's curvature and strength.
Remarkably, this velocity may be expressed also as the curvature of the effective field ${\bf h}_{\text{eff}}(t)$ in the moving frame
\begin{eqnarray}
v_{\overline{\theta }_{1,2}}(t)\equiv \pm [h_{\cal N}(t) \partial_t h_z(t) -h_z(t) \partial_t h_{\cal N}(t) ] \,.
\end{eqnarray} 
Since $|{\bf B}(t)|$ is always positive, one can conclude that ${\bf h}_{\text{eff}}(t)$ has a zero winding around $\hat{\cal T}$. Hence, the velocity of the (instantaneous) fixed points $P_{1,2}(t)$ averaged over a period vanishes.


To proceed further, we linearize Eqs.~(\ref{eq:gyro_spherical}) around
the instantaneous fixed points $P_{1,2}(t)$, which gives the Jacobian
\begin{equation*}
J=%
\begin{pmatrix}
0 & \pm K(t) \\ 
\mp K(t)\ \frac{K(t)^{2}}{\left[ |{\bf B}(t)|^{2}+K(t)^{2}\right] } & 0\ 
\end{pmatrix} \,,
\end{equation*}%
with eigenvalues
\begin{equation*}
E_{J}=\pm i \frac{K(t)^{2}}{\sqrt{|{\bf B}(t)|^{2}+K(t)^{2}}}  \, ,
\end{equation*}%
independently of the positions of the points $P_{1,2}(t)$. Notice
that, if $K(t)\ne 0$, these eigenvales are purely imaginary and
different from zero for any value of the field amplitude and
curvature. This means that, nearby the points $P_{1,2}(t)$, the
trajectories of the spin velocity field in the FSB sphere form closed
loops, namely they have a vortex-like profile at any time position
along the parametric trajectory.


\begin{figure}[t!]
\includegraphics[width=0.97\columnwidth]{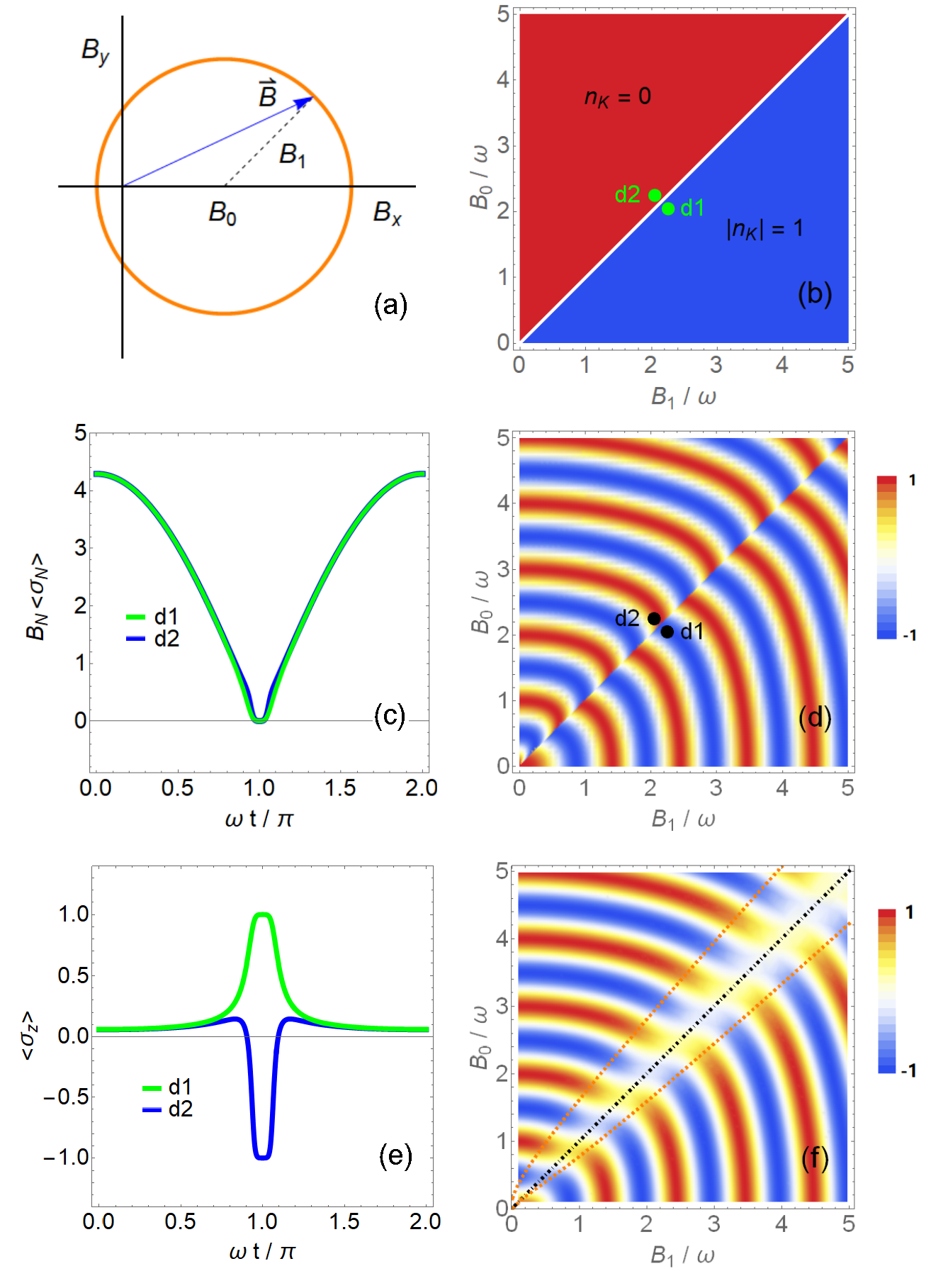}
\caption{(a) Schematic illustration of the shifted circular drive 
in the $(B_x,B_y)$ field components plane. (b) Dependence of the winding number on the field parameters of the shifted circular drive. The diagonal line in the phase diagram (i.e. $B_0=B_1$) corresponds to a topological transition where a jump of the winding number of the applied field occurs. $d_1$ ($d_2$) are two representative points close to the topological transition boundary lying in a domain of the phase diagram with $n_K=1$ ($n_K=0$), respectively. (c) time dependent evolution of the term $|{\bf B}(t)|\langle \sigma_{\cal N} \rangle$ within the AARF solution which appears in the expression of the total quantum phase positions $d_1$ and $d_2$ of the phase diagram. We notice that this contribution is not sensitive to the topological change. (d) contour map of the cosine of the total phase in the parameters space for the near-adiabatic solution. We notice that a sharp dislocation in the wavefront occurs at the topological transition line. (e) time dependent profile of the $z-$ component of the spin within the AARF solution evaluated at $d_1$ and $d_2$. (f) Contour map of the cosine of the total phase obtained from the exact dynamics of the two-level driven system. We notice that the sharp dislocation in (d) is smeared when considering the full dynamics of the spin. In (f) the dotted line (black) indicates the topological line boundary, the red dotted lines include a region where the ratio of the maximal vortex velocity ($v_{\theta}$) with respect to the maximum of the effective field ($|{\bf h}_{\text{eff}}|$) in the rotating frame is larger than one.}
\label{fig:fig3}
\end{figure}   

This is numerically confirmed in Figs.~2(d)-(f), where we show different snapshots of the spin flow in the $(\theta,\varphi)$ plane by depicting the spin vector velocity, represented by an arrow, for different values of the curvature. Fig. 2(d) corresponds to a case of vanishing curvature [point $D$ of the field trajectory in Fig. 2(c)]. In this situation, there is no time gradient in the azimuthal angle, so that the spin velocity is uniform. For a non-vanishing amplitude of the curvature [Figs. 2(e) and 2(f)], the torque can vanish at $\varphi=0,\,\pi$ for values of the azimuthal angles that can be positive or negative depending on the sign of $K(t)$. As expected, the spin velocity flow exhibits a vortex structure around these points.  

A closer look to the vortex structure reveals that, for a large
amplitude of the curvature, the spin velocity flow winds around the
core of the vortex even for values of $\theta$ far from it, so that
about all the points on the sphere (i.e. any spin orientation) are
influenced by the presence of the vortex [Fig. 2(f)]. On the contrary,
for smaller values of $K(t)$, the influence of the vortex on the Bloch
sphere is limited to spin orientation angles close to the positions of
the points with vanishing torque [Fig. 2(e)]. We also notice that the
spin flow always winds in opposite directions around the vortex cores
situated at $\varphi=0$ and $\pi$. Moreover, for a given vortex, the
winding may be clockwise or anticlockwise depending on the value of
the polar angle and the sign of the curvature.

The overall dynamical scenario can be immediately visualized on the
basis of these simple snapshot features. A change in the parametric
space (e.g. time) modifies the spin velocity pattern by rocking the
vortices back and forth from the north (south) pole to the equator in
the $\cal N$-$z$ plane, with a velocity [Eq.~(\ref{eq:speedvortex})]
governed by the field and curvature amplitudes and their
derivatives. During their motion, the vortices expand or shrink
depending on the strength of the curvature. Then, when the system is
prepared in a spin configuration at a given position in the parametric
space, the evolution of spin trajectory is dictated by whether the
spin i) is trapped and pinned by the vortex, ii) succeeds to avoid its
attraction, or iii) is deflected by the vortex path, i.e. its
trajectory is scattered by the vortex motion. In general, for
competing curvature and field strengths, all cases from i) to iii)
cooperate to determine the global spin dynamics.

In conclusion, the dynamical evolution of the spin in the moving frame
is clearly controlled by the presence of two {\it topological} objects
on the Bloch sphere whose motion results from the competition between
the curvature of the applied field and the strength of the field
itself. Each vortex is pinned to move in the $\cal N$-$z$ plane and
generally drives the motion of the spin by modifying the pattern of
the spin velocity flow through the variation of its size on the Bloch
sphere (i.e. via the curvature of the applied field in the rest frame)
and its velocity (i.e. via the curvature of the field in the moving
frame).





\section{Topological imprints on two-level dynamics}


Here, we examine the case of a driving field texture undergoing a topological transition and its effects on the quantum dynamics of a TLS by applying the general results of Sec. III and, especially, the AARF solution for the total phase, Eq. (\ref{NAA-3}). In particular, we intend to isolate the role and consequences of time-dependent field curvatures associated to different topologies. To this aim, we revisit a paradigmatic example involving two coplanar fields: (i) a rotating one with a frequency $\omega$ and amplitude $B_1$ and (ii) a uniform one with amplitude $B_0$, see Fig. \ref{fig:fig3}(a). This configuration was considered recently in Refs. \cite{SVLBNNF15} and \cite{RBSVLNF17}, where imprints of the topological characteristics of the driving field were identified in the quantum phases. Such an effect is attributed (with some degree of approximation) to the windings of the resulting spin textures in the Bloch's sphere.  
 
The model Hamiltonian reads
\begin{eqnarray*}
{\cal H}(t)= (B_0+B_1 \cos \omega t) \sigma_x + B_1 \sin \omega t \sigma_y \,.
\end{eqnarray*}
In the rotating frame this Hamiltonian can be expressed as
\begin{eqnarray*}
{\cal H}(t)= |{\bf B}(t)| \sigma_{\cal N}(t) \, ,
\end{eqnarray*}
with $|{\bf B}(t)|=\sqrt{({B_0}^2+{B_1}^2+2 B_0 B_1 \cos \omega t)}$ the instantaneous magnitude of the total applied field and $\sigma_{\cal N}$ the Pauli matrix associated to the spin projection along $\hat{\cal N}$
\begin{eqnarray*}
\sigma_{\cal N}(t)=[f_x(t) \sigma_x +f_y(t) \sigma_y] \,,
\end{eqnarray*}
where the directors of $\sigma_{\cal N}(t)$ are $f_x(t)=\frac{(B_0+ B_1 \cos \omega t)}{|{\bf B}(t)|}$ and $f_y(t)=\frac{B_1 \sin \omega t}{|{\bf B}(t)|}$ and the corresponding field curvature $K(t)$ can be obtained from Eq. (\ref{Eq:ktgen}). 

As shown in Fig. \ref{fig:fig3}(b), the driving field's winding $n_K$ defined in Eq. (\ref{Eq:windingK}) has a transition along the line $B_0 = B_1$ reflecting a change in the field's topology. The exact solution demonstrates that the topological transition in the driving field leaves a definite imprint on the total phase, Eq. (\ref{eq:totalphase}), in the form of a dislocation along the critical line, as reported in Refs. \cite{SVLBNNF15} and \cite{RBSVLNF17} [see Fig. \ref{fig:fig3} (f)]. A strictly adiabatic treatment would explain this in terms of Berry phases \cite{L-G93}. However, the spin dynamics is far from being adiabatic in the proximities of the critical line. Recent non-adiabatic treatments \cite{SVLBNNF15,RBSVLNF17,BSRF17} have approached the problem in terms of effective Berry phases linked to the winding parity of spin textures. However, in Sec. III we demonstrated that the total phase does not depend explicitly on the winding of the spin texture [Eq.~(\ref{eq:totalphase})], indicating that the field topology can be more relevant than the spin topology in setting the behavior of the quantum phase across the transition. Interestingly, these limitations are overcome by the expression for the total phase obtained within the AARF, Eq. (\ref{NAA-3}), since it explicitly captures the contributions from both the non-adiabatic geometric phases and the field topology, as shown in Fig. \ref{fig:fig3}(d).

Indeed, from the inspection of the total phase [Eq.~(\ref{eq:totalphase})], we observe that the integrand is proportional to that appearing in the dynamical phase [Eq. (\ref{eq:dynamic1})] with an additional factor that depends on $\langle \sigma_z \rangle$. Here, the AARF solution is extremely instructive to understand how the total phase changes when crossing the topological boundary. The key observation is that, within this approximation, the spin orientation is always parallel to the effective field. In particular, the evolution of $\langle \sigma_z\rangle$ is governed by $K(t)$. In Sec. \ref{NAA}, we showed that the field curvature $K(t)$ changes its sign for a non-winding field trajectory, while it keeps a uniform sign if the field non-trivially winds around the origin. As a consequence, $\langle \sigma_z\rangle$ behaves analogously [see Fig. \ref{fig:fig3}(e)]. This behaviour differs from that of $\langle \sigma_{\cal N}\rangle$, governed by $|{\bf B}(t)|$, which has always the same sign on both domains of the phase diagram [see Fig. \ref{fig:fig3}(c)]. The strong dependence of the sign of $\langle \sigma_z\rangle$ on the field's winding determines the ultimate response of the total phase, Eq.~(\ref{eq:totalphase}), to the field's topology. Thus, we find that the AARF reproduces almost every feature of the dislocation pattern in the parameters space [see Fig.~\ref{fig:fig3}(d)], except for the smoothing observed near the topological boundary when the exact dynamics is considered [Fig.~\ref{fig:fig3}(f)].


\begin{figure}[t!]
\includegraphics[width=0.97\columnwidth]{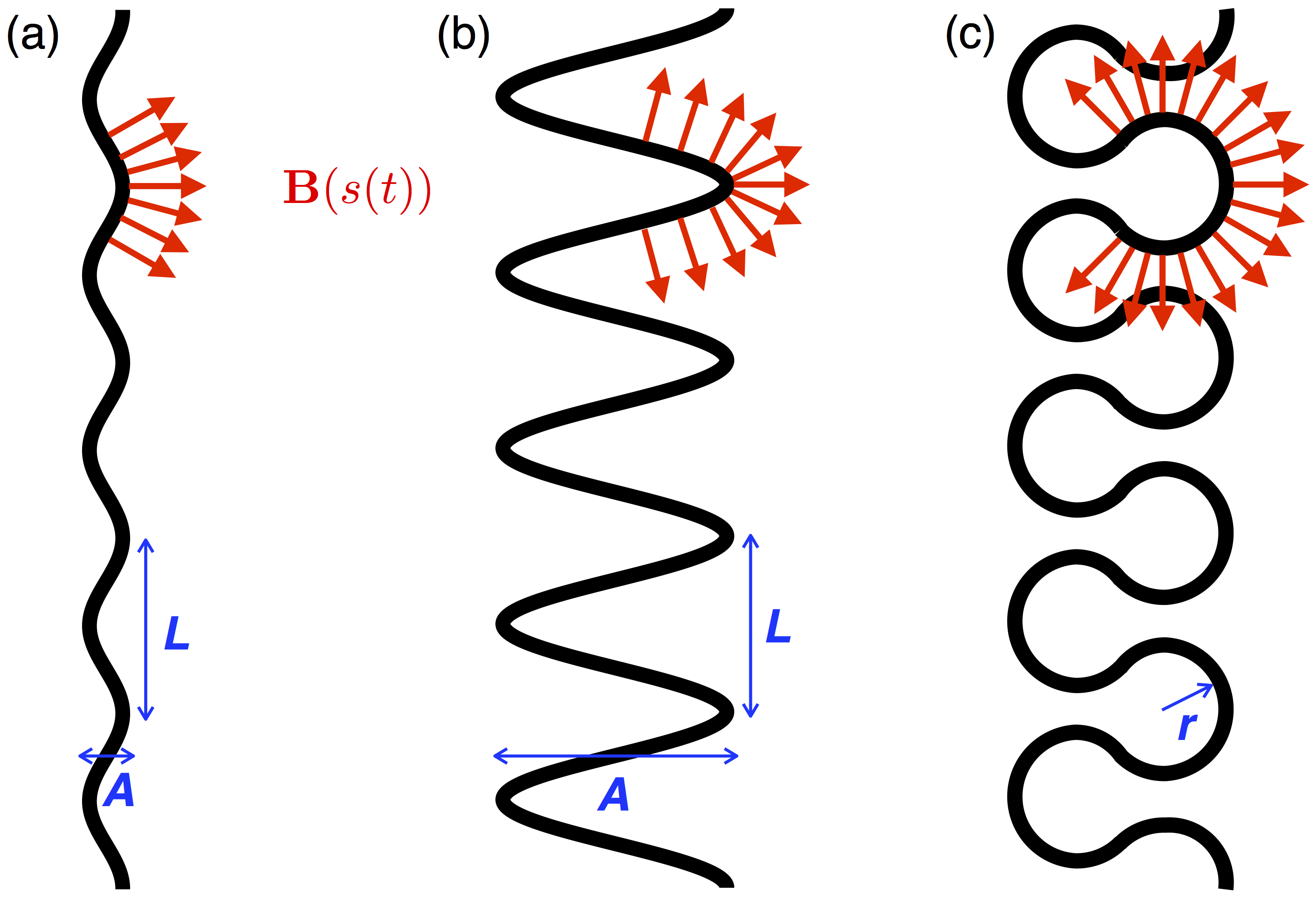}
\caption{1D Rashba spin-orbit channels of varying local curvatures $\kappa[s(t)]$ implementing pendulum-like driving fields ${\bf B}[s(t)]$ of different angular amplitudes $\varphi_0$ and field curvatures $K(s(t))$ according to Eq. (\ref{C3}): (a) $A \ll L$ corresponding to $\varphi_0 < \pi/4$, where elliptical integrals can be simplified; (b) $A > L$ corresponding to $\pi/4 < \varphi_0 < \pi/2$; (c) with $\pi/2 < \varphi_0 < \pi$. The arrows indicate the local orientation of the field ${\bf B}[s(t)]$.}
\label{fig:fig4}
\end{figure}   
 
\begin{figure}[t!]
\includegraphics[width=0.97\columnwidth]{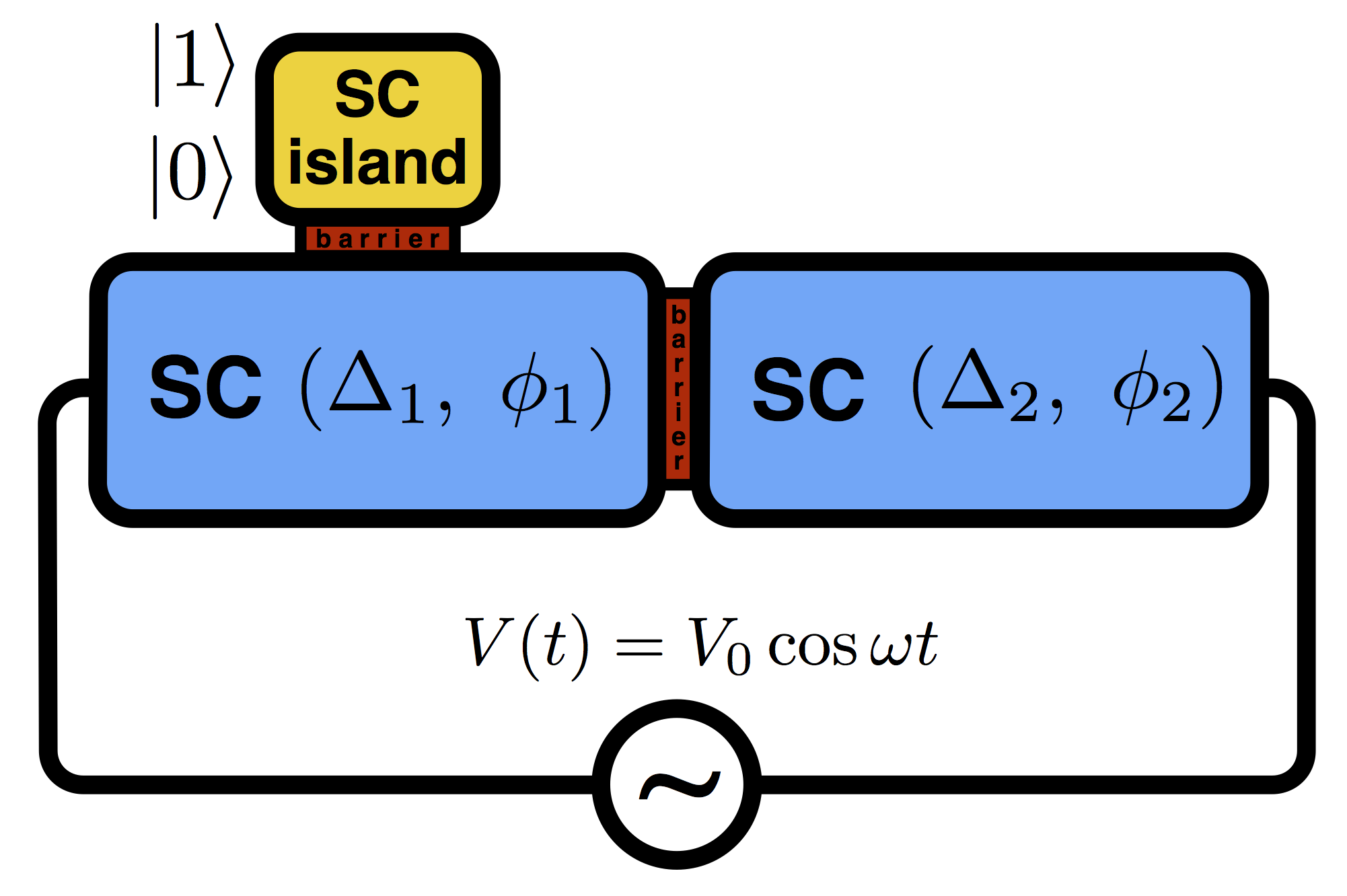}
\caption{Schematic of a superconducting platform for simulating a TLS with pendular drive. The superconducting (SC) island is in an effective Cooper pair box with two relevant states indicated as $|0\rangle$ and $|1\rangle$. The SC island is coupled to a superconductor which is a part of a Josephson junction subjected to an external applied voltage $V(t)=V_0 \cos(\omega t)$. $\Delta_i$ and $\phi_i$ with $i=1,2$ are the amplitude and phase of the order parameter of the superconductors forming the Josephson junction, respectively.}
\label{fig:fig5}
\end{figure}   


\section{Application to different physical platforms}

In this section we discuss a series of quantum platforms where the proposed geometric/topological driving can be experimentally implemented. We focus on pendular-like drivings exploiting geometrical effects due to a changing field curvature despite the trivial topology. The pendular drive is particularly striking. Firstly, from a theoretical point of view, it is a paradigmatic example to highlight the differences between the AA and AARF approximation. Moroever, it can be directly exploited to demonstrate the quantum geometric driving based on the control of the field curvature.
Still, drivings with non-trivial topologies are also considered along the first part of the discussion leading to Eq. (\ref{C3}) and the closing paragraphs.

We start by mapping the Rashba model for a generically shaped (quasi) 1D quantum wire on an effective spin 1/2 system in the presence of a parametric planar driving. We shall demonstrate how the curvature of the wire and the strength of the Rashba interactions build up the amplitude and the curvature of the effective field. To this aim, we follow Refs. \cite{O15,ying16} for the description of spin-orbit coupled electrons on 1D curved space. The corresponding Hamiltonian reads
\begin{equation}
\mathcal{H}=-\frac{\hbar^2}{2m^*} \left[ \partial_s^2 +\frac{\kappa(s)^2}{4} \right]-i \hbar \alpha_{\rm R} \left[ \sigma_{\cal N} \partial_s - \sigma_{\cal T} \frac{\kappa(s)}{2}\right],
\label{Hrashba-0}
\end{equation}
where $s$ is the arclength along the 1D curve, $\kappa(s)$ is the local curvature, and $\alpha_{\rm R}$ is the Rashba coupling strength. In the limit $\hbar|\kappa(s)|/2 \ll |\langle - i \hbar \partial_s \rangle| =  |\langle p_s \rangle| = p_{\rm F}$, with $p_{\rm F}$ the effective Fermi momentum, the Hamiltonian (\ref{Hrashba-0}) reduces to 
\begin{equation}
\mathcal{H}=-\frac{\hbar^2}{2m^*}  \partial_s^2 -i \hbar \frac{\alpha_{\rm R}}{2}  (\sigma_{\cal N} \partial_s +\partial_s \sigma_{\cal N}).
\label{Hrashba-1}
\end{equation}

This approximation corresponds to the semiclassical limit $\lambda_{\rm F} \ll 4\pi r(s)$, with $\lambda_{\rm F}$ the Fermi wavelength and $r(s)= 1/|\kappa(s)|$ the local curvature radius \cite{footnote-1}. The link between spatial and time dependence is straightforward by assuming that the spin carriers propagate along the curve with constant Fermi velocity $v_{\rm F}$, i.e. $\frac{\partial s}{\partial t}=v_{\rm F}$.
Furthermore, by means of a simple algebraic transformation of the Hamiltonian $\mathcal{H}$, we can observe that a spin eigenmode $|\psi(s)\rangle$ of $\mathcal{H}$ evolves in space according to $i \hbar \partial_s |\psi(s)\rangle =\frac{\alpha_{\rm R} m^{*}}{\hbar} \sigma_{\cal N} |\psi(s)\rangle$ \cite{footnote-map}, and, in turn, by introducing the Fermi momentum,
$i \hbar \partial_t |\psi(t)\rangle =\frac{\alpha_{\rm R} p_F}{\hbar} \sigma_{\cal N}|\psi(t)\rangle$. Then, the spin of the carrier, while propagating along the 1D curve, experiences an effective driving field ${\bf B}= \frac{\alpha_{\rm R} p_F}{\hbar} \hat{\cal N}$. 
To make more explicit the correspondence between the space and time pendular field curvature, we notice that by differentiating $\hat{\cal N}$ one finds 
\begin{eqnarray}
\frac{\partial \hat{\cal N}}{\partial t} &=& \frac{\partial \hat{\cal N}}{\partial s} \frac{\partial s}{\partial t} \nonumber \\
&=& \kappa(s) v_{\rm F} \hat{\cal T} 
\label{C1} \\
&=& K(t) \hat{\cal T}, 
\label{C2}
\end{eqnarray}
where we applied the FS-type equation in Eq. (\ref{C1}) \cite{O15,ying16} and Eqs. (\ref{Eq:sigmaN}) and (\ref{Eq:ktgen}) in (\ref{C2}). 
This means that the instantaneous field curvature $K(t)$ is proportional to the local wire curvature $\kappa (s)$ at $s(t)$ and to $v_{\rm F}$, i.e.,
\begin{equation}
K(t)=\kappa[s(t)] v_{\rm F}.
\label{C3}
\end{equation}
This shows that a desired field curvature $K(t)$ can be obtained by designing an appropriate wire curvature $\kappa(s)$ satisfying Eq. (\ref{C3}), unfolding whole families of open and closed curves for curvature-assisted spin interferometry. Moreover, we notice that additional driving-field engineering can be done by introducing Dresselhaus spin-orbit coupling and/or uniform in-plane magnetic fields. Relevant implementations already exist using electrons surfing on surface acoustic waves along winding semiconductor channels \cite{SKGOKNSS13}. A few illustrative examples of 1D quantum wires implementing Rashba pendulum-like drivings of increasing amplitude are depicted in Fig. \ref{fig:fig4}
\cite{footnote-2}.
At this point, an additional remark is required. The analogy between the spatial components of the Rashba field and the pendular driving shows that the possibility of accessing angular amplitudes $\varphi_0$ of order $\pi$ is strongly tied to the shape of the nanostructure. Indeed, as we have schematically showed in Fig.~\ref{fig:fig4}, one needs to modify the profile of the serpentine accordingly to get into dynamical regimes with $\varphi_0$ larger than $\pi/4$. According to our results (Fig. \ref{fig:fig1}), an appropriate choice for the driving field strength (i.e., $\alpha_{\rm R}$ for the Rashba spin-orbit nanochannel) allows to access the regime of geometric driving of the quantum phases whenever $\varphi_0$ is in the range [0,$\pi$].

Another prospective platform to realize a pendular driving can be achieved by means of superconducting materials. We start by considering a small superconducting island in a regime of charge qubit with two relevant states active in the Cooper pair box which correspond to the presence or absence of excess Cooper pairs. Hence, if we assume that the transition between the states $|0\rangle$ and $|1\rangle$ in the island can occur due to a pair tunnelling between the island and another superconductor $S_1$ acting as a reservoir, Fig. \ref{fig:fig5}, the effective low-energy Hamiltonian can be expressed as 
\begin{eqnarray}
\mathcal{H}_{S}=E_{J} (c^{\dagger}_{\uparrow} c^{\dagger}_{\downarrow} \sigma_{-}+ c_{\downarrow} c_{\uparrow} \sigma_{+} ), 
\label{HS1}
\end{eqnarray}
\noindent where $E_{J}$ is the Josephson coupling, the matrices $\sigma_{\pm}$ describe the dynamics in the subspace $\{|0\rangle,|1\rangle\}$ and the operators $c,c^{\dagger}$ are related to the fermionic degree of freedom in the superconductor close to the Fermi level (for convenience of notation we drop the index of the momentum of the Cooper pairs). Then, taking into account that the pairs in $S_1$ are in the condensed ground state, one can replace the fermionic term with the corresponding expectation value associated with the amplitude ($\Delta_1$) and phase ($\phi_1$) of the superconducting order parameter, so that $H_{S}$ reads 
\begin{equation*}
\mathcal{H}_{S}=E_{J} (\Delta_1 \exp[i \phi_1] \sigma_{-}+ \Delta_1 \exp[-i \phi_1] \sigma_{+}) \, .
\end{equation*}
With simple algebraic steps, one can recast $H_{S}$ in the following form
\begin{equation}
\mathcal{H}_{S}=E_{J} \Delta_1 (\cos[\phi_1] \sigma_{x}+ \sin[\phi_1] \sigma_{y}) \, ,
\end{equation}   
\noindent thus corresponding to a TLS with an effective planar field with strength $B=E_{J} \Delta_1$ and whose components are modulated by the phase difference between the island and $S_1$. To complete the building of the time dependent pendular driving, we consider the superconductor $S_1$ as a part of a Josephson junction (Fig. \ref{fig:fig5}) subjected to an external voltage. Taking into account that the basic equation ruling the dynamics of the Josephson effect concerning the phase difference $\phi=(\phi_1-\phi_2)$ across the junction and the applied voltage $V$ is given by $V(t)=\frac{\hbar}{2 e} \frac{\partial \phi}{\partial t}$, with $\frac{\hbar}{2 e}$ being the magnetic flux quantum $\Phi_0$, one can design the phase dynamics in $H_S$ by suitably selecting the time dependence of $V(t)$. Indeed, by means of the harmonic applied voltage $V(t)=V_0 \cos(\omega t)$, we have that the phase $\phi_1(t)$ (less of an offset due to the phase of $S_2$) is oscillating with a frequency $\omega$ that is set by the external electric field and a maximal angular extension of the pendulum $\phi_0=\frac{V_0}{\omega \Phi_0}$. By correspondence of $H_S$ with the Eqs. (\ref{pend-1a}) and (\ref{pend-1b}), we observe that the dynamics of the two levels in the superconducting island (Fig. \ref{fig:fig5}) is very well suited to simulate a TLS pendular drive. 
For completeness, we also notice that the effective low-energy coupling in Eq. (\ref{HS1}) can also emerge in other physical contexts where the existence of the two levels is due to the formation of local (e.g. impurity) electronic states in a metallic host that prefer to be either empty or doubly occupied thus forming pairing centers that in turn can drive superfluid-to-insulator transitions \cite{cuocoBF1,cuocoBF2} or lead to inhomogeneous topological phases \cite{brzezicki}. The fact that such types of pairing centers can have phononic or excitonic origin \cite{PC1}, and can also occur at the surface of topological insulators or Dirac materials \cite{PC2}, indicates that other coherent quantum materials platform with setups similar to that proposed in Fig. \ref{fig:fig5}, but with different drivings, can be also achieved.   

Finally, a literal interpretation of Hamiltonian (\ref{Hgen}) suggests the study of magnetic resonance setups. In nuclear magnetic resonance (NMR), field-curvature effects could be demonstrated experimentally by shaping radio frequency pulses generating suitable driving Hamiltonians in the rotating frame of the nuclear spins \cite{WS88}. 

We point out that field-curvature effects can be significant for the design of shaped pulses for robust quantum control \cite{SA16}. However, the field engineering is very limited in NMR commercial equipments. Another alternative worth to mention is to turn to strongly-driven superconducting qubits (SCQs) \cite{OYLBLO05}, where high-order multiphoton interferometry has been demonstrated and the systems can apparently be easily adapted to a wide spectrum of driving fields and curvatures as proposed here.


\acknowledgments

This work was supported by Project No. FIS2014-53385-P and No. FIS2017-86478-P (MINECO/FEDER, Spain). D.F. acknowledges additional support from the Marie Sk{\l}odowska-Curie Grant Agreement No. 754340 (EU/H2020). 
C.O. acknowledges support from a VIDI grant (Project 680-47-543) financed by the Netherlands Organization for Scientific Research (NWO).
Z.-J.Y. acknowledges support from the National Natural Science Foundation of China (Grant No. 11974151).

%
%

{}


\begin{thebibliography}{}
\bibitem{footnote-0}
Following the derivations explicitly presented in this manuscript, an extension to non-coplanar fields can be straightforwardly achieved.
\bibitem{footnote-1}
Notice that our expression for the semiclassical limit in terms of the local curvature $r(s)$ is uncommon in the literature, where the $\lambda_{\rm F}$ is usually compared to the system's size for defining that limit. We are proposing a stronger condition.
\bibitem{footnote-map}
We notice that the Hamiltonian $\mathcal{H}$ for the Rashba spin-orbit nanochannel can be written as $\mathcal{H}=\tilde{\mathcal{H}}^2-\frac{\alpha_{R}^2}{4 r} \sigma_0$ with $r=\frac{\hbar^2}{2 m^{*}}$ and $\tilde{\mathcal{H}}$ being

\begin{equation}
\tilde{\mathcal{H}}=i \sqrt{r} \partial_s -\frac{\alpha_R}{2} \sqrt{r} \sigma_{\cal N}(s) \,.
\end{equation}


Hence, it is justified to assume that the spin eigenmode evolves according to the linear equation in $\partial_s$ as reported in the main text.
\bibitem{footnote-2}
From Eq. (\ref{C3}) we see that finding the actual wire's shape necessary to reproduce a given field curvature $K(t)$ may require to solve elliptic integrals. For a purely Rashba ring of constant radius $r_0$ the relationship is straightforward, instead: $K= v_{\rm F}/r_0 = 2\pi/T =\omega$, namely, the field curvature $K$ is just the orbital frequency $\omega$.
\end{thebibliography}
\end{document}